\documentclass[10pt,journal,final,onecolumn]{IEEEtran}
\usepackage{xcolor}
\definecolor{mycolor}{RGB}{65,105,225}
\usepackage{graphicx}
\graphicspath{{./Figures/}}

\usepackage{amsmath}
\usepackage{graphicx}

\usepackage{verbatim}
\usepackage{amssymb}
\usepackage{bm}
\usepackage{algorithmic}  
\usepackage{algorithm}  
\usepackage{booktabs}
\usepackage{cite}
\usepackage{amssymb}
\usepackage{nomencl} 
\usepackage{etoolbox}
\usepackage{titlesec}
\titleformat{\section}[block]{\normalfont\Large\bfseries}{\thesection}{1em}{}
\titlespacing*{\section}{0pt}{\baselineskip}{\baselineskip}
\usepackage[colorlinks,linkcolor=black, citecolor=black, urlcolor=black]{hyperref}
\ifCLASSOPTIONcompsoc
    \usepackage[caption=false, font=normalsize, labelfont=sf, textfont=sf]{subfig}
\else
\usepackage[caption=false, font=footnotesize]{subfig}
\fi

\usepackage[normalem]{ulem}
\newcommand\redsout{\bgroup\markoverwith{{\rule[0.5ex]{3pt}{1.2pt}}}\ULon}

\begin{document}
\title{Deep learning based on Transformer architecture for {power system} short-term voltage stability assessment with class imbalance
}

\author{Yang~Li\textsuperscript{a, *}, Jiting~Cao\textsuperscript{b},
Yan~Xu\textsuperscript{c}, Lipeng~Zhu\textsuperscript{d}, 
Zhao~Yang~Dong\textsuperscript{c, *}

\leftline{\textsuperscript{a} School of Electrical Engineering, Northeast Electric Power University, Jilin 132012, China}

\leftline{\textsuperscript{b} State Grid Chengde Power Supply Company, Chengde 067000, China}

\leftline{\textsuperscript{c}  School of Electrical and Electronic Engineering, Nanyang Technological University, 639798, Singapore}

\leftline{\textsuperscript{d}  College of Electrical and Information Engineering, Hunan University, Changsha, China}

\leftline{\textsuperscript{*} Corresponding author. E-mail address: liyang@neepu.edu.cn (Y. Li), zy.dong@ntu.edu.sg (Z.Y. Dong).
}
}

\maketitle

\textbf{Abstract:} Most existing data-driven power system short-term voltage stability assessment (STVSA) approaches presume class-balanced input data. However, in practical applications, the occurrence of short-term voltage instability following a disturbance is minimal, leading to a significant class imbalance problem and a consequent decline in classifier performance. This work proposes a Transformer-based STVSA method to address this challenge. By utilizing the basic Transformer architecture, a stability assessment Transformer (StaaT) is developed {as a classification model to reflect the correlation between the operational states of the system and the resulting stability outcomes}. To combat the negative impact of imbalanced datasets, this work employs a conditional Wasserstein generative adversarial network with gradient penalty (CWGAN-GP)  for synthetic data generation, aiding in the creation of a balanced, representative training set for the classifier. Semi-supervised clustering learning is implemented to enhance clustering quality, addressing the lack of a unified quantitative criterion for short-term voltage stability. {Numerical tests on the IEEE 39-bus test system extensively demonstrate that the proposed method exhibits robust performance under class imbalances up to 100:1 and noisy environments, and maintains consistent effectiveness even with an increased penetration of renewable energy}. Comparative results reveal that the CWGAN-GP generates more balanced datasets than traditional oversampling methods and that the StaaT outperforms other deep learning algorithms. {This study presents a compelling solution for real-world STVSA applications that often face class imbalance and data noise challenges.} 


\textbf{Word count:} 5945.
\makenomenclature
\mbox{}
\section*{Nomenclature}

\textit{Abbreviations}\\
\begin{tabular}{ll}
STVS & Short-term voltage stability \\
STVSA & Short-term voltage stability assessment \\
TDS & Time-domain simulation \\
PMU   & Phasor measurement units \\
DT   & Decision tree \\
TS & Time series \\
ELM & Extreme learning machine  \\
CNN & Convolutional neural network \\
GCN & Graph convolutional network \\
LSTM & Long-short term memory \\
GAN & Generative adversarial network \\
BiGRU & Bi-directional gated recurrent unit \\
StaaT & Stability assessment Transformer  \\
SMOTE & Synthetic minority oversampling technique \\
ROS & Random oversampling \\
ADASYN  & Adaptive synthetic \\
WGAN  & Wasserstein GAN \\
WGAN-GP  & WGAN with gradient penalty \\
CWGAN-GP & Conditional WGAN-GP \\
\end{tabular}
 \newpage
\begin{tabular}{ll}
SFCM & Semi-supervised fuzzy C-means \\
CGAN & Conditional generative adversarial network \\
FC & Fully connected \\
SC & Silhouette coefficient \\
ACC & Accuracy \\
Mis & Misdetection \\
Fal & False-alarm \\
MCC & Matthews correlation coefficient \\
WD & Wasserstein distance \\
MMD & Maximum mean difference \\
FID & Fréchet incidence distance \\
COP-k-means & Constraint-partitioning k-means  \\
OTW & Observation time window \\
SNR & Signal-to-noise ratio \\
\end{tabular}

\vspace{1em}
\noindent\textit{Notations/Symbols}\\
\begin{tabular}{ll}
$X$ & Dataset with all samples \\
$N$ & The dimension of dataset $X$ \\
$m$ & The number of samples in $X$ \\
$C$ & The number of clusters \\
$x_i$ & The $i$th sample in $X$ \\
$S_j$ & Fuzzy membership \\
$J$ & Objective function of SFCM\\
$u_{i j}$ & The extent to which sample $x_i$ is affiliated with category $j$ \\
$c_j$  & The $j$th cluster center \\
$c_s$ & The $s$th cluster center \\
$l$ & Exponent weight \\
$G$ & Generator\\
$D$ & Discriminator\\
$x$ & Real data \\
$z$ & Noise\\
$y$ & Label \\
$\mathcal{D}$ & The set of 1-Lipschitz functions\\
$\widetilde{x}$ & Generated data\\
$\hat{x} $ & Interpolation between $x$ and $\widetilde{x}$\\
$P_{r}$ & Probability distributions followed by real data\\
$P_{z}$ & Probability distributions followed by noise\\
$P_{g}$ & Probability distributions  followed by generated data\\

$P_{\hat{x}}$ & Probability distributions  followed by $\hat{x}$\\
$\gamma$ & Possible joint distributions combined by $P_{r}$  and $P_{g}$\\
$\|f\|_{L}$ & Lipschitz constant of the function $f$\\
$k$ & A constant serving as the upper bound for the $\|f\|_{L}$ \\
$\lambda$ & Gradient penalty coefficient\\

$Q$ & Input feature matrix for queries\\
$K$ & Input feature matrix for keys\\
$V$ & Input feature matrix for values\\
$d_K$ & The dimension of $K$\\
$W^Q$ & Weight matrix for $Q$\\
$W^K$ & Weight matrix for $K$\\
$W^V$ & Weight matrix for $V$\\
$W^O$ & Output weight matrix applied to the concatenated result of all heads\\
$U$ & Voltage amplitude\\ 
$P$ & Active power\\
$Q_r$ & Reactive power\\
$S$ & Time series dataset\\
$n$ & The number of all subsets in dataset $S$\\
$L$ & The number of system buses\\

\end{tabular}
 \newpage
\begin{tabular}{ll}
$S_h$ & The $h$th subset in $S$ \\
$d$ & The dimension of $S_h$\\ 
$f()$ & A continuous function of $x$\\
$V(D,G)$ & Objective function of GAN\\
$L(G)$ & Critical loss of the $G$\\ 
$L(D)$ & Critical loss of the $D$\\
${\left\| {{x_i} - {c_j}} \right\|}$ & Distance from $x_i$ to $c_j$\\

$\mathbb{E}_{x \sim P_r}$ & Expected value of $P_r$\\

$\Pi\left(P_{r}, P_{g}\right)$ & The set of all possible joint distributions combined by $P_{r}$  and $P_{g}$\\
$\inf \limits_{\gamma \in \Pi\left(P_{r}, P_{g}\right)} \mathbb{E}_{\left(x, \widetilde{x}\right) \sim \gamma}\left[\left\|x-\widetilde{x}\right\|\right]$ & Lower bound of the expected value of the distance between samples $x$ and $\widetilde{x}$ \\
$\left\|\cdot\right\|_2$  &  2-norm\\
\end{tabular}

\vspace{1em}
\noindent\textit{Units}\\
\begin{tabular}{ll}
s & Seconds \\
dB & Decibel\\
\% & Percent
\end{tabular}

\section{Introduction}
Short-term voltage stability (STVS) of power systems  signifies their capacity to recover their voltages to an acceptable range following a disturbance, a crucial characteristic linked to the systems' stability\cite{hatziargyriou2020definition}. With the swift escalation in electricity demand coupled with an inadequately expanded transmission system, the power system is often pushed to operate near its stability threshold, which increasingly threatens the secure operation of the system. The problem is further intensified by the escalating prevalence of induction motor loads\cite{potamianakis2006short} and the growing penetration of renewable energy \cite{hasan2019existing}, both of which significantly compound the challenge of maintaining short-term voltage stability in power systems \cite{shair2021power}.

To prevent a severe voltage dip during large disturbances, it is crucial to assess power system STVS. Specifically, the system's ability to maintain short-term voltage stability after a short-term large disturbance needs to be determined. Thus, the short-term voltage stability assessment (STVSA) issue is receiving widespread attention. Early research on STVSA mostly started from the physical mechanism, including the load dynamics of induction motors. Kawabe et al. \cite{kawabe2014analytical} proposed a STVS analysis method based on the P-V plane, assessing the STVS by checking the state of the induction motor. Similarly, a scalable dynamic load model based on bifurcation theory for assessing the STVS in large-scale power systems is proposed in the work of Stankovi{\'c} et al. \cite{stankovic2015dynamic}. The primary approach for model-driven  STVSA predominantly involves time-domain simulation (TDS). However, the short duration of transient processes combined with the large scale of power systems complicates the provision of timely information about STVS through model-driven methods \cite{ge2020model}. Owing to the advent of wide-area measurement systems, phasor measurement units (PMUs) have gained prominence due to their capability to acquire large volumes of high-precision {synchronized measurements}. Consequently, these advancements {have led to the widespread adoption} of data-driven methods for stability assessment \cite{kamwa2009development}. Reference \cite{dasgupta2013real} introduced a model-free STVSA approach utilizing the Lyapunov exponent. However, obtaining reliable results may become challenging when the Lyapunov exponent approximates zero. Utilizing PMU data, shallow machine learning has proven to be beneficial in the area of STVSA. In reference \cite{pinzon2019real}, the authors proposed a machine learning-based method for multi-state STVS classification, demonstrating low classification error and strong predictive capabilities. Further, the work of Zhu et al. \cite{zhu2015time} demonstrated an online STVSA method that amalgamated decision tree (DT) and time series (TS) classification using shapelets. Duchesne et al.  \cite{duchesne2020recent} provided a detailed introduction to the principles of applying machine learning to the stability assessment of power systems. Reference \cite{zhang2018hierarchical} presented a hierarchical adaptive data analysis by using an extreme learning machine (ELM) to conduct real-time STVSA. This ELM-based approach is capable of analyzing not just voltage instability, but also fault-induced delayed voltage recovery. Building upon this, Zhang et al. \cite{zhang2018missing} proposed a real-time STVSA method that can withstand data loss, {enabling STVSA} even in scenarios of PMU data loss. Lastly, the work of Lashgari et al. \cite{lashgari2022fast} proposed a fast DT-based machine learning scheme for timely and accurate prediction of voltage stability and identification of instability drivers.  

With the continuous evolution of deep learning techniques, several deep learning models have been deployed to assess the STVS of power systems. For instance, Hagmar et al. \cite{hagmar2020voltage} developed a supplementary warning system based on long short-term memory (LSTM), leveraging both real-time and historic data to accurately predict voltage instability minutes into the future, while Huang et al. \cite{huang2021distribution} employed a convolutional neural network (CNN) for STVSA. Meanwhile, another study \cite{wang2021short} proposed a STVSA method combining graph convolutional networks (GCNs) and LSTM to capture both temporal and spatial features. In a similar vein, Luo et al. \cite{luo2021data} extracted spatial features via GCN and deployed one-dimensional convolution networks to learn temporal information. Furthermore, the work of Zhu et al. \cite{zhu2020intelligent} employed LSTM to determine the STVS of power systems and proposed intelligent assessment schemes. Finally, reference \cite{li2022deep} adopted a generative adversarial network (GAN) to enhance the sample and assess STVS utilizing a bi-directional gated recurrent unit (BiGRU). However, LSTMs are not designed for parallel computing and tend to struggle with processing long sequence data. In contrast, CNNs can handle parallel computation effectively, but they are {less capable when it comes to managing long-distance dependencies within the data}. The inherent limitations of these algorithms prevent them from satisfying the requirements for the rapid processing of features that span long distances in data sequences. This need is accentuated by the increasing complexity of power systems, thereby compromising the assessment performance significantly \cite{zhang2022utl}. Within the field of computer vision, the Transformer model, as highlighted in reference \cite{vaswani2017attention}, disrupted the traditional reliance of deep learning on CNNs and recurrent neural networks (RNNs), and in doing so, established an encoder-decoder model exclusively grounded in self-attention mechanisms. {Beyond this,} the Transformer has found wide-ranging applications, including but not limited to load forecasting \cite{gao2023adaptive} and  forced oscillation localization in power systems \cite{matar2023transformer}. Consequently, this study simplifies the Transformer, incorporates it into STVSA, and {puts forth} the stability assessment Transformer (StaaT).

{In existing research 
on STVSA in power systems, the substantial amount of required data is typically generated by simulations \cite{li2022deep,zhong2022method}.} During this process, the errors of the used model and its parameters are inevitable. Enhancing the reliability of the assessment model becomes significantly feasible when the measured data is procured from actual PMUs. However, the real post-disturbance system tends to be predominantly stable, resulting in a serious imbalance in the  stable-to-unstable ratio. If not addressed appropriately, {this sort of class imbalance problem can severely undermine} the performance of the model\cite{zhang2021deep}. In the STVSA field, reference \cite{zhu2017imbalance} utilized a hybrid method that combines oversampling with cost-sensitive learning. This strategy not only considers the degree of imbalance but also weighs the cost of misclassification. 
Nonetheless, cost-sensitive learning necessitates the use of expert knowledge for meticulously establishing the cost matrix. Given the escalating complexity of application problems and the potential constraints stemming from experts' experience and knowledge, the accuracy of the cost matrix may be compromised. In general, {data resampling serves as the most prevalent} approach to achieving a balanced sample distribution \cite{zhu2017imbalance}. Data resampling techniques primarily include oversampling, undersampling, and mixed sampling. Classic oversampling techniques such as synthetic minority oversampling technique (SMOTE), random oversampling (ROS), and adaptive synthetic (ADASYN) sampling have been widely used. {However, these techniques are not without their limitations.} For example, linear oversampling methods like SMOTE and ROS are prone to overfitting, while ADASYN is susceptible to outliers. The GAN, proposed in reference \cite{goodfellow2014generative}, is extensively utilized for data resampling, primarily due to the authenticity and diversity of its generated samples.  However, conventional GANs grapple with training difficulties and the need for careful coordination of generator and discriminator parameters. To mitigate these challenges, the Wasserstein GAN (WGAN) was introduced in reference \cite{arjovsky2017wasserstein}. Subsequently, reference \cite{gulrajani2017improved} proposed the Wasserstein GAN with gradient penalty (WGAN-GP) to prevent gradient vanishing in WGAN. The current study utilizes a conditional WGAN-GP (CWGAN-GP) to tackle the class imbalance problem in STVSA. This marks a significant advancement over existing methods with known drawbacks, such as the propensity for overfitting in SMOTE and ROS, and sensitivity to outliers in ADASYN. The proposed CWGAN-GP not only addresses these limitations but also optimizes the neural network's learning capacity by ensuring uniform weight distribution. The method relies on balanced datasets generated by CWGAN-GP for model training, {thereby circumventing the deterioration of} deep learning model performance caused by class imbalance during training.

To {underscore the limitations of existing methodologies} and the novelty of this study, {a comparison and analysis of the most recent} research in the field of data-driven STVSA {has been conducted}. The {findings} are {presented} in Table \ref{tab1}.

\begin{table}[h]\centering
\caption{Comparison of Current Study with Previous Research in Data-Driven STVSA}
	\label{tab1}
\setlength{\tabcolsep}{6mm}
\begin{tabular}{ccccc}
\hline\hline
Reference  & \begin{tabular}[c]{@{}c@{}}Class  imbalance\end{tabular} & \begin{tabular}[c]{@{}c@{}}Imbalanced learning method\end{tabular}     & \begin{tabular}[c]{@{}c@{}}Deep  learning\end{tabular} & \begin{tabular}[c]{@{}c@{}}Self-attention  mechanism\end{tabular} \\ \hline
\cite{zhu2020time}    & —                                                          & —                                                                         & —                                                        & —                                                                   \\
\cite{zhang2018hierarchical}   & —                                                          & —                                                                         & —                                                        & —                                                                   \\
\cite{zhang2018missing}   & —                                                          & —                                                                         & —                                                        & —                                                                   \\
\cite{huang2021distribution}   & —                                                          & —                                                                         & $\surd$                                                        & —  \\
\cite{wang2021short}   & —                                                          & —                                                                         & $\surd$                                                        & —                                                                   \\
\cite{luo2021data}   & —                                                          & —                                                                         & $\surd$                                                        & —                                                                   \\
\cite{zhu2020intelligent}   & —                                                          & —                                                                         & $\surd$                                                        & —                                                                   \\
\cite{zhu2017imbalance}   & $\surd$                                                          & \begin{tabular}[c]{@{}c@{}}oversampling, cost-sensitive learning\end{tabular}        & —                                                        & —                                                                   \\
\cite{ren2019fully}  & —                                                          & —                                                                         & $\surd$                                                        & —                                                                   \\
This paper & $\surd$                                                          & \begin{tabular}[c]{@{}c@{}}generative adversarial network\end{tabular} & $\surd$                                                        & $\surd$                                                                   \\ \hline\hline
\end{tabular}
\end{table}

Considering all the issues discussed, a STVSA method is proposed based on CWGAN-GP and StaaT in  environments with class imbalance. This work's principal contributions include:
\begin{enumerate}
    \item The imbalance learning method based on CWGAN-GP is proposed to address the class imbalance problem in STVSA. Unlike the oversampling method that generates samples by simple linear sampling, this technique  generates a balanced, {realistic} dataset from real data. Moreover, the method proves resilient under class imbalances up to 100:1, beyond which its effectiveness decreases, underscoring its practicality in dealing with real-world data imbalance challenges. 
    \item  This study designs a STVSA model based on stability assessment Transformer. This model leverages its multi-head self-attention mechanism to enhance efficiency through parallel computing, enabling it to simultaneously learn the characteristics of different positions in the time series samples.
    {\item An additional contribution of this study lies in demonstrating that the proposed method maintains robust performance even in scenarios with increased penetration of renewable energy. The exploration of how renewable energy integration impacts the STVSA enriches the understanding of complex post-disturbance dynamics and stability boundaries in power systems, thereby enhancing the method's practical applicability.}
\end{enumerate}

The suggested approach undergoes testing on the IEEE 39-bus test system under class imbalance conditions. In the test, various statistical indexes were selected for testing. Experimental results indicate a superior performance of the proposed method over alternatives. Moreover, this study conducts a sensitivity analysis on the parameters that may affect the results.

\section{ Related algorithms}\label{sec.related_work}
\subsection{Semi-supervised Clustering Algorithm}\label{sfcm}
{ A significant challenge within the realm of STVSA} is the absence of a uniform quantitative criterion, making the process of acquiring accurate sample labels complex. The consecutive labeling of all training samples, when relying solely on domain knowledge, can potentially lead to inefficiencies. However, depending exclusively on unsupervised learning with unlabeled data may yield inconsistent results. Fortunately, the well-understood concept of voltage stability can offer crucial insights. For instance, {as outlined by }reference\cite{zhu2015time}, if none of the bus voltages drop below 0.9 pu for a period of 10 seconds, the system can be confidently categorized as stable; in contrast, a definitive label of instability can be assigned if all bus voltages drop below 0.7 pu without recovery. This clear understanding {allows for the accurate labeling} of a small subset of samples, {which can subsequently serve as prior information} for clustering the remaining unlabeled samples. Therefore, this study employs a semi-supervised clustering algorithm to enhance the sample labeling process.

Based on these clearly labeled samples, the semi-supervised fuzzy C-means (SFCM) method can be used to judge the stability of each sample. An \emph{N}-dimensional dataset $X={x_i} (1 \leqslant i \leqslant m)$ is divided into $C$ clusters, and each sample belongs to clusters through fuzzy membership $S_j (1 \leqslant j \leqslant C)$. In semi-supervised learning, these labeled samples in $X$ function as prior knowledge. The SFCM algorithm {leverages this information from labeled data to guide} the process of iterative optimization of the overall clustering. The objective function $J$ of SFCM is 
\begin{equation}\label{eq.J}
 J = \sum\limits_{i = 1}^m {\sum\limits_{j = 1}^C {u_{ij}^l{{\left\| {{x_i} - {c_j}} \right\|}^2}{\rm{ }}} }  
\end{equation}
where $u_{i j}$ {indicates the extent to which} sample $x_i$ {is affiliated with} category $j$; $l$ denotes the exponent weight; $c_j$ is the $j$th cluster center; and $\left\| {{x_i} - {c_j}} \right\|$ represents the distance from $x_i$ to $c_j$. The specific representations of the degree of membership $u_{ij}$ and the cluster center $c_j$ are
\begin{equation}
u_{i j}=\frac{1}{\sum_{s=1}^{C}\left(\frac{\left\|x_{i}-c_{j}\right\|}{\left\|x_{i}-c_{s}\right\|}\right)^{\frac{1} {l-1}}}
\end{equation}
\begin{equation}
c_{j}=\frac{\sum_{i=1}^{m} u_{i j}^{l} \cdot x_{i}}{\sum_{i=1}^{m} u_{i j}^{l}}
\end{equation}
where $c_s$ refers to the $s$th cluster center. Due to space constraints, the specific steps of the SFCM will not be elaborated further. For more details on SFCM, please refer to references  \cite{mai2015semi,li2022deep}.

\subsection{Conditional Wasserstein GAN with Gradient Penalty}
In recent years, GANs have played a crucial role in data resampling, generating renewable scenarios \cite{li2021privacy}, and addressing renewable output uncertainties in economic dispatch problems\cite{li2022hierarchical}. The generator's objective is to produce an ample number of denotes samples, while the discriminator's mission is to distinguish between generated and real samples with the highest possible accuracy. Through this dynamic competition, they mutually constrain and evolve. As a result, the generated samples more closely resemble the real ones, thus obtaining the desired samples. The objective function $V(D,G)$ of GAN is 
\begin{equation}
 \begin{aligned}
\min _{G} \max _{D} V(D, G)=\mathbb{E}_{x \sim P_r}[\log D(x)]+\mathbb{E}_{z \sim P_z}[\log (1-D(G(z)))]
\end{aligned}
\end{equation}
where $G$ denotes the generator, while $D$ stands for the discriminator. Noise is denoted by $z$, whereas $P_r$ and $P_z$ are the probability distributions followed by real data $x$ and noise $z$ respectively. The expected value is given by $\mathbb{E}_{x \sim P_r}$. Random noise is fed into the generator to generate synthetic data, which is then input into the discriminator alongside the real data. The discriminator's {job is to} determine the authenticity of the input. When real samples are inputted, the result inclines towards 1, while with fake samples, it leans towards 0. Through training, when either real or generated samples are fed into the system, the outcome of $V(D,G)$ is 0.5, indicating that the generator can, {at this point, create convincingly} real pseudo-samples. 

The conditional GAN (CGAN) , an extension of GAN \cite{mirza2014conditional}, has the following objective function: 
\begin{equation}\label{eq.V-CGAN}
 \begin{aligned}
 \min _{G} \max _{D} V(D, G)=\mathbb{E}_{x \sim P_r}[\log D(x \mid y)]+\mathbb{E}_{z \sim P_z}[\log (1-D(G(z) \mid y))]
  \end{aligned}
\end{equation}
where $y$ can be any type of additional information. 

However, the original GAN has inadequacies in training and needs to carefully coordinate the training level of $G$ and $D$ \cite{li2023data}. To address this concern, WGAN was proposed in \cite{arjovsky2017wasserstein}, and an approach to removing log() from the objective functions of $G$ and $D$ was suggested, expressed as 
\begin{equation}
\min _{G} \max _{D \in \mathcal{D}} V(D, G)=\mathbb{E}_{x \sim P_r}[D(x)]\\+\mathbb{E}_{z \sim P_z}[1-D(G(z))]
\end{equation}
where $\mathcal{D}$ represents the collection of 1-Lipschitz functions. By using the $D$ of WGAN to fit distance instead of classification, the discriminator of the original GAN performs a dual-classification task. Thus, the sigmoid {function} of the last layer of $D$ is removed. The study also indicates that if two distributions do not overlap or the overlap is negligible in a high-dimensional space, Kullback–Leibler divergence and Jensen–Shannon divergence can neither reflect the distance nor provide the gradient. The recommended metric for providing a meaningful gradient is the original Wasserstein distance, defined as follows:
\begin{equation}
W\left(P_{r}, P_{g}\right)=\inf _{\gamma \in \Pi\left(P_{r}, P_{g}\right)} \mathbb{E}_{\left(x, \widetilde{x}\right) \sim \gamma}\left[\left\|x-\widetilde{x}\right\|\right]
\end{equation}
where $P_{g}$  denotes the probability distributions of generated data; $\Pi\left(P_{r}, P_{g}\right)$  is the set of all possible joint distributions combined by $P_{r}$  and $P_{g}$. For each possible joint distribution $\gamma$ ,  real data $x$ and generated data $\widetilde{x}$ can be derived from $\left(x, \widetilde{x}\right) \sim \gamma$ . The lower bound $\inf \limits_{\gamma \in \Pi\left(P_{r}, P_{g}\right)} \mathbb{E}_{\left(x, \widetilde{x}\right) \sim \gamma}\left[\left\|x-\widetilde{x}\right\|\right]$  of the expected value of the sample distance in all possible joint distributions is defined as Wasserstein distance. As $\inf \limits_{\gamma \in \Pi\left(P_{r}, P_{g}\right)}$  cannot be solved directly, it was transformed into \cite{arjovsky2017wasserstein}: 
\begin{equation}
W\left(P_{r}, P_{g}\right)=\frac{1}{k} \sup _{\|f\|_{L} \leq k} \mathbb{E}_{x \sim P_{r}}[f(x)]-\mathbb{E}_{x \sim P_{g}}[f(x)]
\end{equation}
where $f(x)$ is a continuous function of $x$, $\|f\|_{L}$ denotes the Lipschitz constant of the function $f$, and the function $f$ is referred to as a $k$-Lipschitz function if it satisfies  $\|f\|_{L} \leqslant k$ for some constant $k$. To satisfy $\|f\|_{L} \leqslant k$ , each time the parameters of $D$ are updated in WGAN, their absolute values are truncated to no more than a fixed constant. However, reference \cite{gulrajani2017improved} suggested that in WGAN, if the threshold $c$, a key parameter constraining the discriminator's weights, is not fine-tuned, issues like gradient disappearance or explosion may occur. The critical losses of the $G$ and $D$ of WGAN-GP are given by \cite{gulrajani2017improved}:
\begin{equation}
 \begin{aligned}
 \left.L(G)=\mathbb{E}_{\widetilde{x} \sim P_g}[D(\widetilde{x})\right]
 \end{aligned}
\end{equation}

\begin{equation}
 \begin{aligned}
 \left.L(D)=\mathbb{E}_{\widetilde{x} \sim P_g}[D(\widetilde{x})]-\mathbb{E}_{x \sim P_r}[D(x)]+\lambda \mathbb{E}_{\hat{x} \sim P_{\hat{x}}}\left[\left\|\nabla_{\hat{x}} D(\hat{x})\right\|_{2}-1\right]^{2}\right]
 \end{aligned}
\end{equation}
where  $\hat{x}$ is an interpolation between $x$ and $\widetilde{x}$, $P_{\hat{x}}$ represents the probability distributions followed by $\hat{x}$, $\left\|\cdot\right\|_2$ denotes the 2-norm, and $\lambda$ is the gradient penalty coefficient. The gradient penalty is independently applied to each sample to ensure that $G(z)$ is close to $x$, and the approaching process $D(G(z))$ does not exceed $D(x)$. In CWGAN-GP, the objective function can be defined as follows:
\begin{equation}\label{eq.V-CWGANGP}
 \begin{aligned}
 \min _{G} \max _{D} V(D, G)=\mathbb{E}_{x \sim P_{r}}[D(x \mid y)]-\mathbb{E}_{\widetilde{x} \sim P_{g}}[D(\widetilde{x} \mid y)]-\lambda \mathbb{E}_{\hat{x} \sim P_{\hat{x}}}\left[\left(\left\|\nabla_{\hat{x}} D(\hat{x} \mid y)\right\|_{2}-1\right)^{2}\right]
 \end{aligned}
\end{equation}
\subsection{Transformer}
The Transformer{, along with its enhanced variants, fundamentally} represents a sequence-to-sequence architecture. This is primarily composed of two significant components: the encoder and the decoder. Within the encoder, multiple identical layers are stacked, each of which includes the multi-head self-attention mechanism and the feed-forward neural network (FNN). The decoder enriches the original design by introducing an encoder-decoder attention sublayer on top of the multi-head self-attention and FNN of the encoder. Further, every sub-layer is complemented with a residual connection and layer normalization. To {preserve the input order for the model}, the Transformer {incorporates} a position code at the input layer.

The self-attention {mechanism treats the sequence by substituting each element with the weighted mean of the remainder of} the sequence. {While attention mechanisms have found application across diverse domains,} reference \cite{vaswani2017attention} introduced the concept of “Scaled Dot-Product Attention”.  The complete formula is {written as} 
\begin{equation}
    \text{Attention}(Q,K,V)= \text{softmax}(\frac{QK^T}{\sqrt{d_K}})V
\end{equation}
where $Q$, $K$, and $V$ are the input feature matrices for {queries}, keys, and values, respectively, while $d_K$ {denotes} the dimension of the {keys} matrix, $K$. Reference \cite{vaswani2017attention} {further proposed the idea of} multi-head attention, {which essentially} processes input features from distinct positions through different weights to derive the final result. The formula is as follows:
\begin{equation}
\begin{gathered}
\text {MultiHead}(Q, K, V)=\text {Concat}\left(\text {head}_{1}, \ldots, \text {head}_{h}\right) W^{O} \\
\text {where head}_{i}=\text {Attention}\left(Q W^{Q}_{i}, K W^{K}_{i}, V W^{V}_{i}\right)
\end{gathered}
\end{equation}
where $W^Q$, $W^K$, and $W^V$ are the weight matrices for query, key, and value, respectively, and $W^O$ is the {output weight matrix applied to the concatenated result of all heads}. The multi-head attention mechanism empowers the Transformer model to glean information from diverse locations{, thus offering an advantage over} the single attention mechanism.

\section{Class imbalance learning based on CWGAN-GP}
\subsection{Model structure}
\begin{figure}
    \centering
    \includegraphics[width=3.4in]{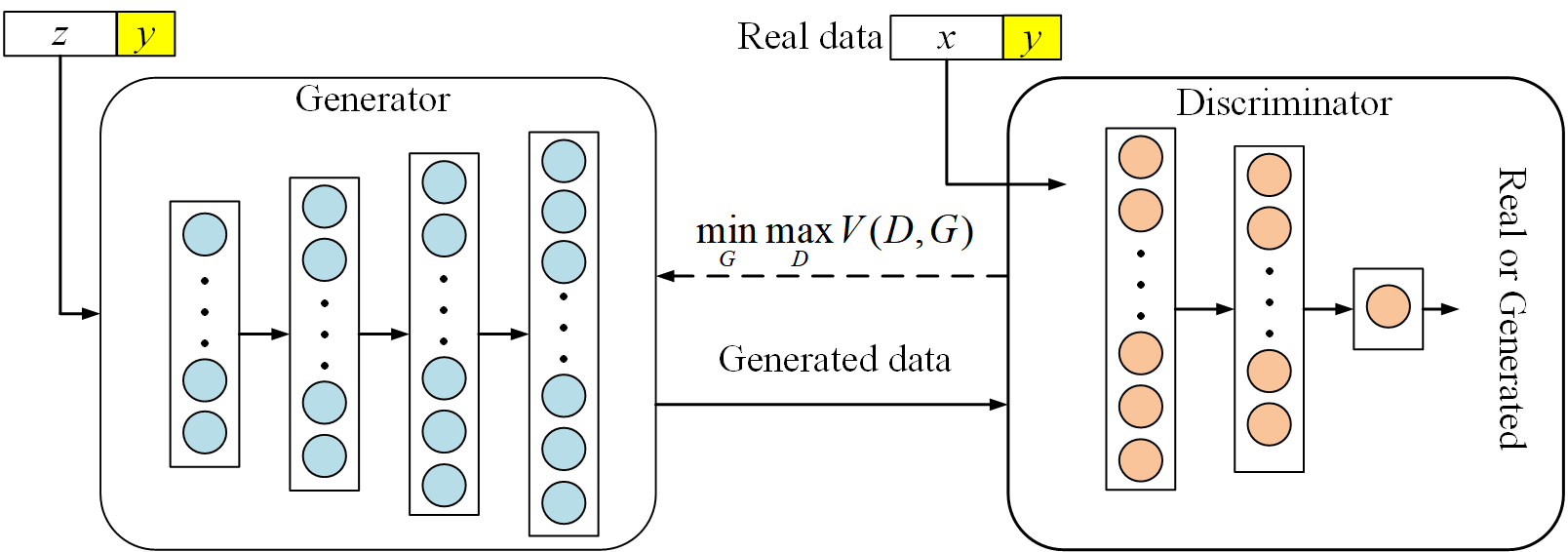}
    \caption{Structure of CWGAN-GP.}\label{fig.2}
    \label{fig:my_label}
\end{figure}

In this study, the CWGAN-GP structure encompasses a generator and a discriminator{, as illustrated in  Fig. \ref{fig.2}}.  
The generator is constructed as a 4-layer fully connected (FC) network. Meanwhile, the discriminator is designed as a three-layer FC network, with each layer comprising 512, 256, and 1 neuron, respectively. The activation function utilized in both the generator and discriminator is LeakyReLU. To train the classifier, the generated samples are employed as the training set. These samples{, generated by the CWGAN-GP model, serve as input data} to train the classifier{, allowing it to learn and generalize from the generated data}.

\subsection{Model training process}
The {model training process of CWGAN-GP} is divided into three steps.

Step 1: Random noise, {denoted as $z$,} and labels, {represented by $y$,} are fed into the generator. The generator is trained to produce generated samples that align with the distribution of real samples.

Step 2: The generated {data} and real {data} are randomly mixed together and input into the discriminator. The discriminator then {evaluates the probability of each sample being real or fake}.

Step 3: Through repeated iterations and optimization, a trained generator is obtained. This trained generator {is capable of generating} a balanced training set{, which can be further utilized for STVSA}.

\section{Proposed STVSA model}

\subsection{Stability assessment Transformer}
\begin{figure}
    \centering
    \includegraphics[width=2.5in]{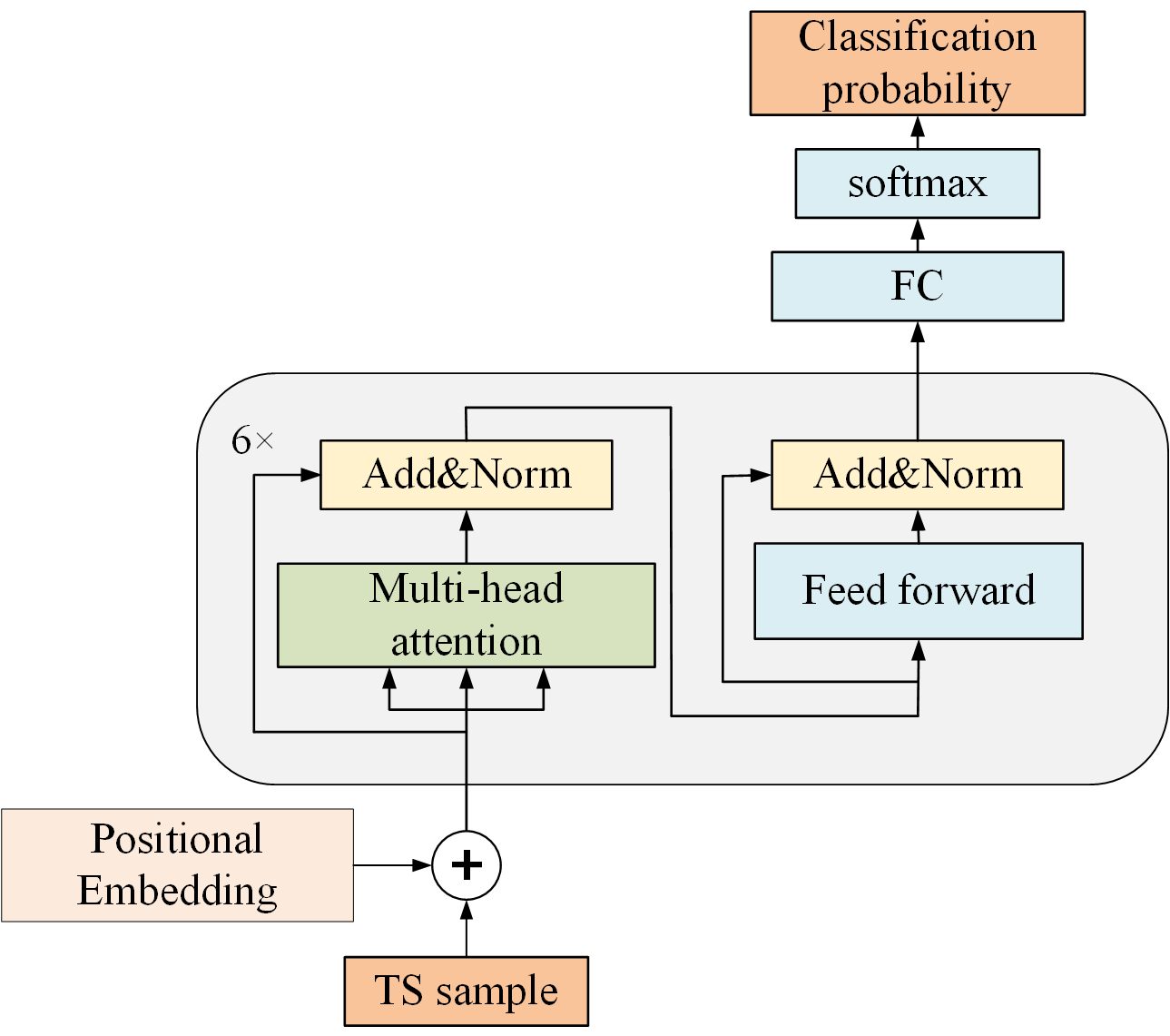}
    \caption{Structure diagram of StaaT.}\label{fig.3}
    \label{fig:my_label}
\end{figure}

When Transformer is applied in natural language processing, the encoder layer is responsible for learning data features, {while} the decoder layer {utilizes these} learned features for prediction. However, in the specific case of a classification task, the focus lies solely on learning the data features. {Consequently}, this study {modifies the original Transformer architecture by removing the decoder layer. Instead, it directly connects} the FC layer and Softmax function after multiple encoder layers to {generate the final model results}. This simplified version of Transformer {is specifically adapted for STVSA purposes}. The structure diagram of StaaT is illustrated in Fig. \ref{fig.3}.

The input to StaaT {consists of} a TS sample comprising real-time power system data collected by PMU{s}. The positional embedding technique is {employed to encode} the data's positional information before it is forwarded to the encoder. {Within} each layer of the encoder, the multi-head attention mechanism {facilitates the learning of data features} through self-attention. {The learned features} are then propagated back through the FNN. The final layer of the encoder outputs the results through a FC layer {followed by a Softmax function, determining} whether the sample is classified as stable or not.
\subsection{STVSA processes}
\begin{figure*}
    \centering
    \includegraphics[width=0.80\textwidth]{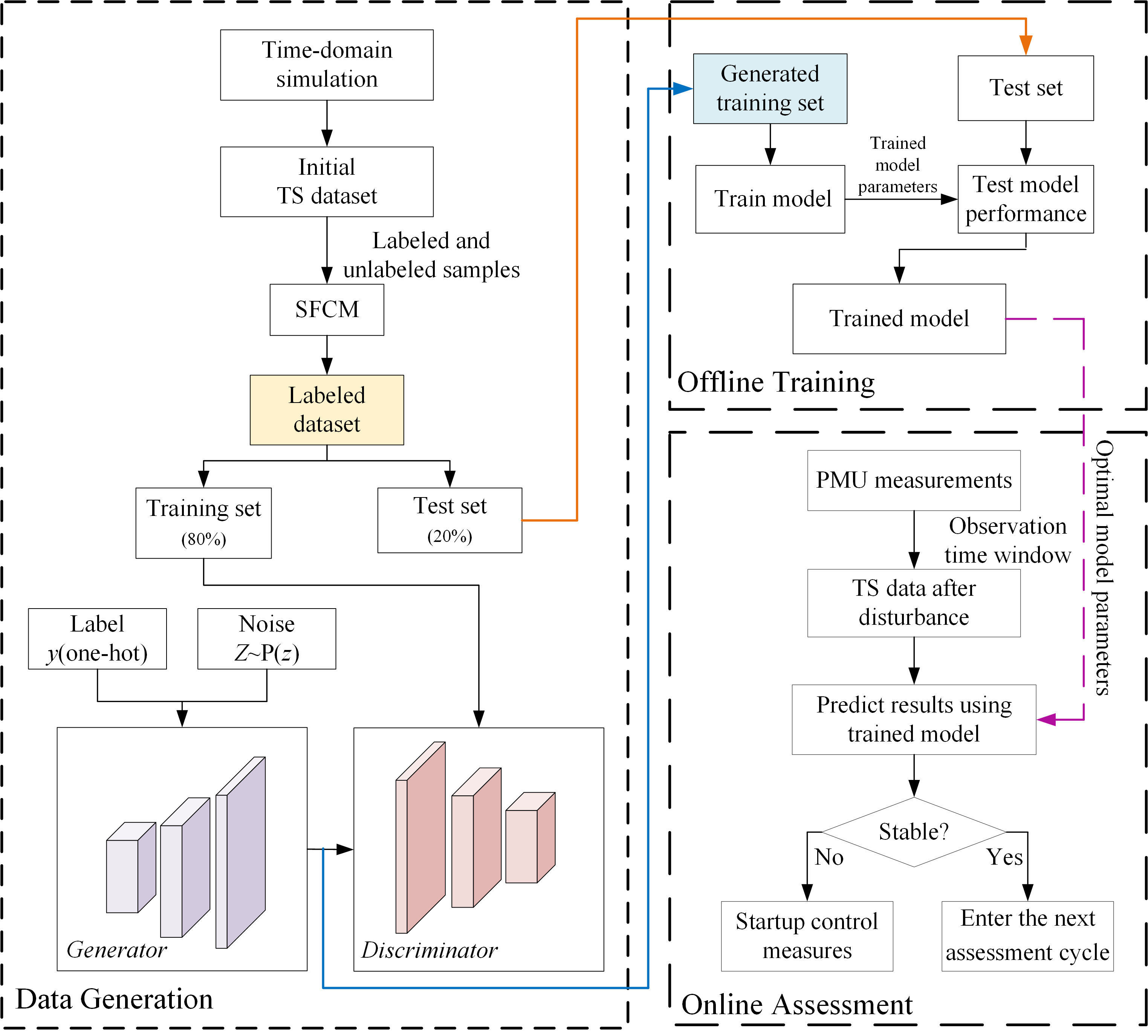}
    \caption{Flowchart of the proposed method.}\label{fig.4}
    \label{fig:my_label}
\end{figure*}

The proposed STVSA scheme in this study consists of three stages, as depicted in  Fig.
\ref{fig.4}.

\textbf{First stage}{:} \emph{Data generation}. Considering different operating conditions in the time-domain simulation, the initial TS samples comprising normalized voltage amplitude ($U$), active power ($P$), and reactive power ($Q_r$) of each bus within a short time period after fault clearance can be obtained \cite{luo2021data}. Subsequently, the unlabeled ones in the initial sample dataset are labeled using SFCM, resulting in a fully labeled dataset. This dataset is subsequently partitioned into training and test sets using a 4:1 ratio \cite{zhu2015time}, with the training set serving as input for the CWGAN-GP model, which generates a balanced training set.

\textbf{Second stage}{:} \emph{Offline training}. The generated training set is utilized to train the StaaT model, while the test set is employed to evaluate its performance. The optimal parameters obtained from the trained model are saved for subsequent online assessments.

\textbf{Third stage}{:} \emph{Online assessment}. Once a big disturbance occurs in the power system, electrical quantities $U$, $P$, and $Q_r$ will be captured in real-time by PMUs and transmitted to the trained StaaT model for assessing the system's stability outcomes. If the system is classified as stable, the monitoring process continues. However, if the system is identified as unstable, appropriate control measures are promptly initiated to mitigate the impact of short-term voltage instability.

\subsection{Assessment indicators}
\subsubsection{Silhouette coefficient}
This study uses silhouette coefficient ($SC$) as an index of the clustering algorithm \cite{li2022deep}, defined as 
\begin{equation}
SC=\frac{1}{N} \sum_{j=1}^{N} \frac{b_{j}-a_{j}}{\max \left(a_{j}, b_{j}\right)}
\end{equation}
where $x_j (1\leqslant j\leqslant N)$ is a sample in the dataset with $N$ samples; $a_j$ denotes the average distance of sample $x_j$ from all other samples in the same cluster, reflecting the degree of aggregation of this cluster; $b_j$ denotes the minimum average distance from $x_j$ to all other clusters, reflecting the degree of separation between different clusters.
\subsubsection{Statistical indexes for classification}
{In the context of} data-driven STVSA, {the misclassification of an unstable sample} as stable can precipitate serious consequences, such as voltage collapse, {due to the system's failure to} initiate the necessary control measures. {Given the varied} costs associated with misclassification, {the reliance solely} on accuracy (ACC) \cite{li2022deep,chicco2020advantages} {proves inadequate} for comprehensive STVSA. It is crucial to employ additional statistical tests, such as the F1-score \cite{chicco2020advantages,li2022deep}, misdetection (Mis) \cite{kamwa2009development,zhu2017imbalance}, false-alarm (Fal)\cite{kamwa2009development,zhu2017imbalance}, and G-mean \cite{zhu2017imbalance}, to comprehensively evaluate the performance. By considering these additional statistical tests, a more comprehensive assessment of the STVSA model's performance can be achieved, ensuring effective detection of unstable samples and minimizing the risks associated with misjudgment.
\subsubsection{Evaluation indicators of GAN}
This paper {adopts} Wasserstein distance (WD) \cite{li2022deep}, maximum mean difference (MMD) \cite{xu2018empirical}, and Fréchet incidence distance (FID) \cite{tran2021data} as quantitative evaluation metrics to assess the performance of CWGAN-GP. A lower value for each of these metrics indicates better model performance. The precise definitions of these three indicators can be found in the references {mentioned earlier}.

\section{Case study}
The tests in this study are performed on the IEEE 39-bus test system, a well-known system in STVSA \cite{li2022deep,ren2019fully,van2020test}, as depicted in Fig. 4. The generator in the system is equipped with a third-order exciter model, and further details can be found in \cite{bu2015psd}.
The dataset is obtained using commercial simulation software, PSD-BPA, which effectively enables accurate simulations and facilitates the acquisition of data for the experiments.
The implementation of the proposed approach is achieved by using PyTorch 1.7 and Tensorflow 1.14 within the Python 3.7 environment. The developed programs are deployed on a personal computer with the following specifications: Intel Core i5-6300HQ 2.3GHz CPU, 4GB RAM, and GTX 960M GPU. These hardware and software configurations facilitate efficient execution of the algorithm and enable comprehensive analysis of the results.
\begin{figure}
    \centering
    \includegraphics[width=3 in]{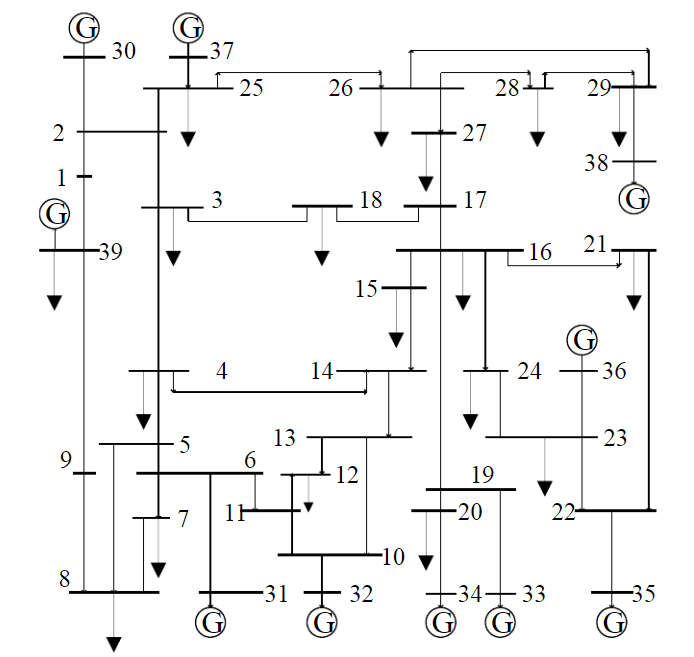}
    \caption{One-line diagram of IEEE 39-bus test system}\label{fig.5}
    \label{fig:my_label}
\end{figure}

The main hyperparameters of CWGAN-GP and StaaT are presented in Table \ref{tab+1} and Table \ref{tab+2} respectively. These hyperparameters have been selected via trial and error. In Table \ref{tab+1}, the parameter $n_{critic}$ is used to strike a balance between the convergence speed of CWGAN-GP training.

\begin{table}[h]\centering
\caption{Hyperparaments setting of CWGAN-GP}
	\label{tab+1}
\setlength{\tabcolsep}{7mm}
\begin{tabular}{ll}
\hline \hline
Hyperparameters         & Values \\ \hline
Gradient penalty factor & 10     \\
Optimizer               & Adam   \\
Learning rate           & 0.0001 \\
Batch size              & 64     \\
$n_{critic}$            & 5      \\
Epoch                   & 500    \\ \hline \hline
\end{tabular}
\end{table}

\begin{table}[h]\centering
\caption{Hyperparaments setting of StaaT}
	\label{tab+2}
\setlength{\tabcolsep}{5mm}
\begin{tabular}{ll}
\hline \hline
Hyperparameters                   & Values \\ \hline
The number of multi-head attention & 8      \\
Optimizer                          & Adam   \\
Dropout                            & 0.5    \\
Learning rate                      & 0.0001 \\
Batch size                         & 64     \\
Epoch                              & 200    \\ \hline \hline
\end{tabular}
\end{table}

\subsection{Dataset generation}
Short-term voltage instability{, which often occurs} following a fault, is predominantly linked to the rapid recovery of loads, especially {those associated with} induction motors \cite{kim2020understanding,zhang2022load}. Hence, in the simulations, the load model of the power system is set up with induction motor load and static ZIP load \cite{li2023pmu}. As illustrated in Table \ref{tab2}, various operating conditions, induction motor load ratios, fault types, and fault locations are designated to simulate crucial operating conditions and emergencies. 

\begin{table}[h]\centering
\caption{Simulation Parameters for Dataset Generation}
	\label{tab2}
\begin{tabular}{ll}
\hline\hline
Type                       & Value                                                                                     \\ \hline
Load level                & 80\%, 100\%, 120\%                                                                             \\
Induction motor load ratio & 70\%, 80\%, 90\%                                                                               \\
Fault type                 & three-phase short circuit                                                                  \\
Fault location             & \begin{tabular}[c]{@{}l@{}}0\%, 25\%, 50\%, and 75\% of the transmission lines\end{tabular} \\
Fault clearing time        & \begin{tabular}[c]{@{}l@{}}0.05 s at the near end; 0.1 s at the far  end\end{tabular} \\ \hline\hline
\end{tabular}
\end{table}

With the configurations set forth in Table \ref{tab2}, real-time system data encompassing voltage amplitude, along with active and reactive power (symbolized as $U$, $P$, and $Q_r$) as per the study \cite{luo2021data}, are procured through time-domain simulation via PSD-BPA, thereby constructing the TS dataset $S$.
 
\begin{equation}
S = [{S_1},{S_2}, \cdots ,{S_h}, \cdots ,{S_n}](1 \le h \le n)
\end{equation}
where $n$ indicates that the dataset $S$ consists of $n$ subsets, and the specific composition is as follows:
\begin{equation}
\begin{aligned}
{S_h} = {[({U_1}, \cdots ,{U_L}),({P_1}, \cdots ,{P_L}),({Q_{r,1}}, \cdots ,{Q_{r,L}})]_1}, \cdots ,\\
{[({U_1}, \cdots ,{U_L}),({P_1}, \cdots ,{P_L}),({Q_{r,1}}, \cdots ,{Q_{r,L}})]_d} 
\end{aligned}
\end{equation}
where $L$ represents the number of system buses, $d$ is the dimension of $S_h$. 

According to the settings in Table \ref{tab2}, a total of 5500 samples are generated in this paper, of which unstable samples account for 9.09\% of the total samples.

\subsection{Performance test of SFCM}
{According to} the acceptance rules outlined in Section \ref{sfcm}, labels can be acquired for all data samples {only in} a minority of cases. The SFCM method incorporates a small subset of data samples with known labels as prior information and {uses it within the SFCM objective function to perform clustering}. Consequently, labels are derived {for all data}. Two distinct label sets are differentiated using constraint-partitioning k-means (COP-k-means) and the engineering criteria mentioned in \cite{zhu2015time}. To assess the efficacy of the SFCM method, the silhouette coefficients of these three label sets are calculated {and used as} performance indicators. The results are presented in Table \ref{tab3}.

\begin{table}[h]\centering
\caption{Comparison of clustering effectiveness}
	\label{tab3}
\begin{tabular}{cccc}
\hline\hline
Index                  & SFCM   & COP-k-means & Engineering criterion \\ \hline
$SC$ & 0.4874 & 0.3954      & 0.2682                \\ \hline\hline
\end{tabular}
\end{table}

{The SFCM method} produces satisfactory results in labeling samples, as shown in Table \ref{tab3}. It demonstrates {nearly} double the efficiency compared to using the engineering criterion, and also exceeds the performance of the COP-k-means algorithm. {These results highlight} the superior efficiency of SFCM in clustering data samples.

\subsection{Performance tests of CWGAN-GP}
To assess the effectiveness of CWGAN-GP in dealing with class imbalance issues, {this study employs several} different data resampling algorithms, including SMOTE, ROS, and ADASYN, on the {same} initial dataset. The specific parameters for these methods are {detailed} in \cite{chawla2002smote}, \cite{batista2004study}, and \cite{alhudhaif2021novel}, respectively. The {assessment} results obtained from StaaT are {summarized} in Table \ref{tab4}.

\begin{table}[h]\centering
\caption{Performance Comparison of CWGAN-GP and Data Resampling Methods}
	\label{tab4}
\scalebox{1}{
\begin{tabular}{cccccc}
\hline\hline
\begin{tabular}[c]{@{}c@{}}Data resampling \\ method \end{tabular}  & ACC(\%)        & MCC             & Mis(\%)        & Fal(\%)           &  G-mean  \\ \hline
ROS                  & 98.87          & 0.9288          & 0.09          & 1.18    & 0.9885      \\
SMOTE                    & 98.91          & 0.9381          & 0.09          & 1.00  & 0.9895          \\
ADASYN                 & 99.09          & 0.9469          & 0.18          & 0.73  &   0.9860          \\
\textbf{CWGAN-GP}      & \textbf{99.82} & \textbf{0.9892} & \textbf{0} & \textbf{0.18} & \textbf{0.9990} \\ \hline\hline
\end{tabular}}
\end{table}

Table \ref{tab4} reveals that CWGAN-GP outperforms the data
resampling algorithms, exhibiting superior performance across all indices. Particularly, the proposed method yields a zero misdetection rate, aligning more with engineering requirements. Furthermore, as alluded to earlier, the MCC offers a clearer representation of algorithm performance in resolving class imbalance. The MCC of CWGAN-GP notably exceeds those of other algorithms by a minimum of 0.04, suggesting that the data generated by CWGAN-GP can better address the issue.

To validate the data generation capabilities of CWGAN-GP, this study conducts a comparison with CGAN in terms of WD, MMD, and FID. The results of this comparison are depicted in Table \ref{tab+3}.

\begin{table}[h]\centering
\caption{Comparison of Data Generation Capabilities: CWGAN-GP vs CGAN}
	\label{tab+3}
\setlength{\tabcolsep}{7mm}
\begin{tabular}{ccc}
\hline \hline
Indicators & CWGAN-GP & CGAN  \\ \hline
WD         & 2.687    & 4.362 \\
MMD        & 0.069    & 0.091 \\
FID        & 1.448    & 2.276 \\ \hline \hline
\end{tabular}
\end{table}

As illustrated in Table \ref{tab+3}, the WD, MMD, and FID metrics associated with CWGAN-GP surpass those of the CGAN. Notably, the WD for CWGAN-GP is 38.4\% lower than that of CGAN, suggesting that CWGAN-GP is superior to CGAN in terms of data generation capabilities.

Aside from algorithmic differences, the proportion between positive and negative samples in the initial dataset can also influence the classifier's performance in addressing class imbalance issues. To investigate this, the proposed method is applied to process and evaluate the performance of the initial dataset with various proportions. The findings are presented in Table \ref{tab5}.

\begin{table}[h]\centering
\caption{Classifier Performance {across Various Positive to Negative Sample Ratios}}
	\label{tab5}
\setlength{\tabcolsep}{4mm}
\begin{tabular}{cccc}
\hline\hline
Proportions & ACC(\%) & MCC    & G-mean     \\ \hline
5:1         & 99.83   & 0.9940 & 0.9970   \\
10:1        & 99.82   & 0.9892 & 0.9990   \\
50:1        & 99.61   & 0.8980 & 0.9477   \\
{100:1}       & {99.60}   & {0.7980} & {0.8935}   \\
{200:1}        & {99.55}   & {0.5453} & {0.7734}   \\
\hline\hline
\end{tabular}
\end{table}

All proposed methods, despite varying ratios of positive to negative samples, yield high-performing samples as demonstrated in Table \ref{tab5}. Although the overall performance remains commendable when the ratio of positive to negative samples in the initial dataset increases, there is a slight decrease in performance. This demonstrates the robust adaptability of CWGAN-GP to varying ratios between positive and negative samples. {Moreover, as the degree of category imbalance escalates, the performance of the proposed method initially remains relatively stable. However, when the degree of imbalance reaches 100:1, the efficiency of the investigated approach. begins to degrade to an unacceptable level. Note that the  metric ACC doesn't significantly fluctuate with increasing imbalance, which suggests it's not entirely reasonable to gauge the effectiveness of the suggested method solely based on accuracy.}
\subsection{Performance test {at different observation times}}
The rapid nature of short-term voltage fluctuations necessitates immediate attention following any significant disruption. If data from shorter observation periods after a major disturbance could be utilized to accurately determine system stability, it would create more time for subsequent protective actions within the power system, thus ensuring better stability. To explore this concept, four different lengths for observation time windows (OTWs) are selected, and their respective accuracies are computed. The findings of this examination are compiled in Table \ref{tab6}.
\begin{table}[h]\centering
\caption{Accuracies at different observation times}
	\label{tab6}
\begin{tabular}{ccc}
\hline\hline
 Lengths of OTW (s) & Training accuracy (\%) & Testing accuracy (\%) \\ \hline
0.03                                                                              & 99.88              & 99.82             \\
0.04                                                                              & 99.81              & 99.78             \\
0.05                                                                              & 99.79              & 99.73             \\
0.06                                                                              & 99.89              & 99.82             \\ \hline\hline
\end{tabular}
\end{table}

As illustrated in Table \ref{tab6}, the model presents high accuracy when the length of OTW is 0.03 s. As this length increases, both the training and testing accuracies remain relatively stable. Given the balance between assessment speed and accuracy, a 0.03s OTW is adopted across all test examples.
\subsection{Performance test of StaaT}
In order to assess the performance of StaaT, {the model is evaluated using a confusion matrix, as presented} in Table \ref{tab7}.

\begin{table}[h]\centering
\caption{Confusion matrix}
	\label{tab7}
\begin{tabular}{ccc}
\hline\hline
                     & Stable (Actual) & Unstable (Actual) \\ \hline
Stable (Predicted)   & 998             & 0                  \\
Unstable (Predicted) & 2                & 100               \\ \hline\hline
\end{tabular}
\end{table}

As demonstrated in Table \ref{tab7}, the StaaT registers a false alarm rate of merely 0.18\%, and a misdetection rate hits 0\%. These outcomes ensure the system's superior accuracy in instability assessments, aligning well with practical requirements. Consequently, the proposed method delivers dependable results and bears substantial practical relevance.

For a more comprehensive evaluation of StaaT, this study compares its test results with other prevalent classification models, namely LSTM, CNN, and BiGRU. The comparison encompasses four metrics: ACC, MCC, F1-score, and Mis. These benchmark algorithms also undergo parameter optimization via a trial and error approach. The summarised outcomes of these comparisons are presented in Table \ref{tab8}.

\begin{table}[h]\centering
\caption{Performance Comparison of StaaT with Other Classification Models}
	\label{tab8}
\begin{tabular}{ccccc}
\hline\hline
Classification model & ACC(\%)       & MCC        & F1-score             & Mis(\%)             \\ \hline
LSTM                 & 95.18          & 0.7179          & 0.9734          & 2.09         \\
CNN                  & 95.82          & 0.7565          & 0.9769          & 1.73          \\
BiGRU                  & 98.55          & 0.9171          & 0.9920          & 0.27          \\
\textbf{StaaT}       & \textbf{99.82} & \textbf{0.9892} & \textbf{0.9990} & \textbf{0} \\ \hline\hline
\end{tabular}
\end{table}

As exhibited in Table \ref{tab8}, all the performance indicators of StaaT outshine those of the contrasted deep learning algorithms. Specifically, boasting the highest MCC and the smallest Mis, StaaT demonstrates superior performance in an environment of class imbalance. This is primarily attributed to the self-attentive mechanism of the StaaT model that allows for comprehensive learning of information at each data position, thereby rendering it more effective than the other alternatives such as CNN, LSTM, and BiGRU.

\subsection{{Performance test with renewable energy integration}}
As renewable energy units continue to merge into the power grid, the dynamic characteristics of the system following disturbances grow increasingly complex \cite{borivcic2021fundamental}. In order to assess the implications of renewable energy unit integration on the proposed methodology, this work examine its performance under varying rates of renewable energy penetration. The outcomes of these assessments are depicted in Table XII.
\begin{table}[h]\centering\caption{Performances of the proposed approach under different renewable energy penetration rates}
	\label{tab13}
\begin{tabular}{cccccc}

\hline\hline
Penetrations rates (\%) & ACC(\%) & MCC    & Mis(\%) & Fal(\%) & G-mean \\ \hline
0                      & 99.82  & 0.9892 & 0      & 0.18   & 0.9990 \\
10                     & 99.73  & 0.9836 & 0.09   & 0.18   & 0.9940 \\
20                     & 99.55  & 0.9726 & 0.18   & 0.27   & 0.9885 \\
30                     & 99.36  & 0.9617 & 0.27   & 0.36   & 0.9829 \\ \hline\hline
\end{tabular}
\end{table}

{Based on the assessment outcomes presented in Table XII, it is evident that the performance of the proposed approach declines in proportion to the increasing penetration of renewable energy sources. This decrease can be attributed to the escalating complexity in post-disturbance dynamics as renewable energy sources are integrated into the power grid, making them more challenging to learn. Nonetheless, the proposed model sustains a commendable performance level even at a penetration rate as high as 30\%. This strongly suggests that the proposed model is capable of delivering effective STVSA for power systems linked to renewable energy sources.}

\subsection{Robustness test under noisy environments}
To assess the robustness of the StaaT model in noisy conditions, Gaussian white noises with varying signal-to-noise ratios (SNRs) are incorporated into the test dataset. However, in real-world scenarios, measurement noise is inevitable during the real-time sampling process conducted by PMUs. The performance under these test conditions is presented in Table \ref{tab11}. 

\begin{table}[h]\centering
\caption{Performance comparison under different SNRs}
	\label{tab11}
\begin{tabular}{cccccc}

\hline\hline
SNR (dB)   & ACC (\%) & MCC    & Fal (\%) & Mis (\%) & G-mean \\ \hline
Noise-free & 99.82   & 0.9892 & 0.18    & 0       & 0.9990 \\
50         & 99.73   & 0.9836 & 0.18    & 0.09    & 0.9940 \\
40         & 99.45   & 0.9678 & 0.45    & 0.09    & 0.9925 \\
30         & 99.18   & 0.9527 & 0.73    & 0.09    & 0.9910 \\ \hline\hline
\end{tabular}
\end{table}

 Table \ref{tab11} demonstrates that, even under various SNR noise conditions, the proposed classifier maintains superior performance. Notably, an increase in the noise proportion does not significantly affect the classifier's performance. This stability under noise disturbance demonstrates the robustness of the proposed method.

\section{Conclusion}

To address the challenge of class imbalance in short-term voltage stability assessment, a novel approach has been developed in this study by leveraging the capabilities of CWGAN-GP and StaaT. The research findings are summarized as follows:

\begin{enumerate}

\item The proposed imbalance learning method based on CWGAN-GP effectively addresses the class imbalance problem in STVSA. The method generates a balanced, realistic dataset from real data and exhibits resilience under class imbalances up to 100:1. The superior performance of the CWGAN-GP is confirmed when compared with CGAN and several data resampling techniques, including SMOTE, ROS, and ADASYN.

\item The employment of StaaT for STVSA enables the extraction of the most important features in the dataset. Notably, it outperforms other alternative methods, demonstrating increased accuracy, improved Matthews Correlation Coefficient, superior F1-score, and minimized misclassification rates.

\item The proposed method demonstrates robust performance even under varying ratios of positive to negative samples and and noisy environments. Moreover, it maintains consistent effectiveness even in scenarios with increased penetration of renewable energy, providing valuable insights into the understanding of complex post-disturbance dynamics in power systems.
\end{enumerate}

Future work will consider real-world scenarios where missing PMU data may occur and extend the testing of the proposed method to larger power system models. Furthermore, combining the data-driven approach of this study with model-driven techniques might offer a comprehensive insight into the causes of voltage instability, thereby facilitating the development of effective corrective measures.

\section*{Acknowledgement}
This work is supported by the Natural Science Foundation of Jilin Province, China under Grant No. YDZJ202101ZYTS149.

\normalem
\bibliographystyle{IEEEtran}
\bibliography{resubmit.bbl}

\begin{thebibliography}{10}
\providecommand{\url}[1]{#1}
\csname url@samestyle\endcsname
\providecommand{\newblock}{\relax}
\providecommand{\bibinfo}[2]{#2}
\providecommand{\BIBentrySTDinterwordspacing}{\spaceskip=0pt\relax}
\providecommand{\BIBentryALTinterwordstretchfactor}{4}
\providecommand{\BIBentryALTinterwordspacing}{\spaceskip=\fontdimen2\font plus
\BIBentryALTinterwordstretchfactor\fontdimen3\font minus
  \fontdimen4\font\relax}
\providecommand{\BIBforeignlanguage}[2]{{%
\expandafter\ifx\csname l@#1\endcsname\relax
\typeout{** WARNING: IEEEtran.bst: No hyphenation pattern has been}%
\typeout{** loaded for the language `#1'. Using the pattern for}%
\typeout{** the default language instead.}%
\else
\language=\csname l@#1\endcsname
\fi
#2}}
\providecommand{\BIBdecl}{\relax}
\BIBdecl

\bibitem{hatziargyriou2020definition}
N.~Hatziargyriou, J.~Milanovic, C.~Rahmann, V.~Ajjarapu, C.~Canizares,
  I.~Erlich, D.~Hill, I.~Hiskens, I.~Kamwa, B.~Pal \emph{et~al.}, ``Definition
  and classification of power system stability--revisited \& extended,''
  \emph{IEEE Transactions on Power Systems}, vol.~36, no.~4, pp. 3271--3281,
  2020.

\bibitem{potamianakis2006short}
E.~G. Potamianakis and C.~D. Vournas, ``Short-term voltage instability: effects
  on synchronous and induction machines,'' \emph{IEEE Transactions on power
  systems}, vol.~21, no.~2, pp. 791--798, 2006.

\bibitem{hasan2019existing}
K.~N. Hasan, R.~Preece, and J.~V. Milanovi{\'c}, ``Existing approaches and
  trends in uncertainty modelling and probabilistic stability analysis of power
  systems with renewable generation,'' \emph{Renewable and Sustainable Energy
  Reviews}, vol. 101, pp. 168--180, 2019.

\bibitem{shair2021power}
J.~Shair, H.~Li, J.~Hu, and X.~Xie, ``Power system stability issues,
  classifications and research prospects in the context of high-penetration of
  renewables and power electronics,'' \emph{Renewable and Sustainable Energy
  Reviews}, vol. 145, p. 111111, 2021.

\bibitem{kawabe2014analytical}
K.~Kawabe and K.~Tanaka, ``Analytical method for short-term voltage stability
  using the stability boundary in the \uppercase{PV} plane,'' \emph{IEEE
  Transactions on Power Systems}, vol.~29, no.~6, pp. 3041--3047, 2014.

\bibitem{stankovic2015dynamic}
A.~M. Stankovi{\'c} and A.~T. Sari{\'c}, ``Dynamic voltage stability assessment
  in large power systems with topology control actions,'' \emph{IEEE
  Transactions on Power Systems}, vol.~31, no.~4, pp. 2892--2902, 2015.

\bibitem{ge2020model}
H.~Ge, Q.~Guo, H.~Sun, and W.~Zhao, ``A model and data hybrid-driven short-term
  voltage stability real-time monitoring method,'' \emph{International Journal
  of Electrical Power \& Energy Systems}, vol. 114, p. 105373, 2020.

\bibitem{kamwa2009development}
I.~Kamwa, S.~Samantaray, and G.~Joos, ``Development of rule-based classifiers
  for rapid stability assessment of wide-area post-disturbance records,''
  \emph{IEEE Transactions on Power Systems}, vol.~24, no.~1, pp. 258--270,
  2009.

\bibitem{dasgupta2013real}
S.~Dasgupta, M.~Paramasivam, U.~Vaidya, and V.~Ajjarapu, ``Real-time monitoring
  of short-term voltage stability using \uppercase{PMU} data,'' \emph{IEEE
  Transactions on Power Systems}, vol.~28, no.~4, pp. 3702--3711, 2013.

\bibitem{pinzon2019real}
J.~D. Pinz{\'o}n and D.~G. Colom{\'e}, ``Real-time multi-state classification
  of short-term voltage stability based on multivariate time series machine
  learning,'' \emph{International Journal of Electrical Power \& Energy
  Systems}, vol. 108, pp. 402--414, 2019.

\bibitem{zhu2015time}
L.~Zhu, C.~Lu, and Y.~Sun, ``Time series shapelet classification based online
  short-term voltage stability assessment,'' \emph{IEEE Transactions on Power
  Systems}, vol.~31, no.~2, pp. 1430--1439, 2015.

\bibitem{duchesne2020recent}
L.~Duchesne, E.~Karangelos, and L.~Wehenkel, ``Recent developments in machine
  learning for energy systems reliability management,'' \emph{Proceedings of
  the IEEE}, vol. 108, no.~9, pp. 1656--1676, 2020.

\bibitem{zhang2018hierarchical}
Y.~Zhang, Y.~Xu, Z.~Y. Dong, and R.~Zhang, ``A hierarchical self-adaptive
  data-analytics method for real-time power system short-term voltage stability
  assessment,'' \emph{IEEE Transactions on Industrial Informatics}, vol.~15,
  no.~1, pp. 74--84, 2018.

\bibitem{zhang2018missing}
Y.~Zhang, Y.~Xu, R.~Zhang, and Z.~Y. Dong, ``A missing-data tolerant method for
  data-driven short-term voltage stability assessment of power systems,''
  \emph{IEEE Transactions on Smart Grid}, vol.~10, no.~5, pp. 5663--5674, 2018.

\bibitem{lashgari2022fast}
M.~Lashgari and S.~M. Shahrtash, ``Fast online decision tree-based scheme for
  predicting transient and short-term voltage stability status and determining
  driving force of instability,'' \emph{International Journal of Electrical
  Power \& Energy Systems}, vol. 137, p. 107738, 2022.

\bibitem{hagmar2020voltage}
H.~Hagmar, L.~Tong, R.~Eriksson \emph{et~al.}, ``Voltage instability prediction
  using a deep recurrent neural network,'' \emph{IEEE Transactions on Power
  Systems}, vol.~36, no.~1, pp. 17--27, 2021.

\bibitem{huang2021distribution}
W.~Huang, W.~Zheng, and D.~J. Hill, ``Distribution network reconfiguration for
  short-term voltage stability enhancement: An efficient deep learning
  approach,'' \emph{IEEE Transactions on Smart Grid}, vol.~12, no.~6, pp.
  5385--5395, 2021.

\bibitem{wang2021short}
G.~Wang, Z.~Zhang, Z.~Bian, and Z.~Xu, ``A short-term voltage stability online
  prediction method based on graph convolutional networks and long short-term
  memory networks,'' \emph{International Journal of Electrical Power \& Energy
  Systems}, vol. 127, p. 106647, 2021.

\bibitem{luo2021data}
Y.~Luo, C.~Lu, L.~Zhu, and J.~Song, ``Data-driven short-term voltage stability
  assessment based on spatial-temporal graph convolutional network,''
  \emph{International Journal of Electrical Power \& Energy Systems}, vol. 130,
  p. 106753, 2021.

\bibitem{zhu2020intelligent}
L.~Zhu, D.~J. Hill, and C.~Lu, ``Intelligent short-term voltage stability
  assessment via spatial attention rectified \uppercase{RNN} learning,''
  \emph{IEEE Transactions on Industrial Informatics}, vol.~17, no.~10, pp.
  7005--7016, 2020.

\bibitem{li2022deep}
Y.~Li, M.~Zhang, and C.~Chen, ``A deep-learning intelligent system
  incorporating data augmentation for short-term voltage stability assessment
  of power systems,'' \emph{Applied Energy}, vol. 308, p. 118347, 2022.

\bibitem{zhang2022utl}
R.~Zhang, W.~Yao, Z.~Shi, X.~Ai, Y.~Tang, and J.~Wen, ``Towards multi-scenario
  power system stability analysis: An unsupervised transfer learning method
  combining {DGAT} and data augmentation,'' \emph{IEEE Trans. Power Syst.},
  early access, Nov. 8, 2022. doi: 10.1109/TPWRS.2022.3220569.

\bibitem{vaswani2017attention}
A.~Vaswani, N.~Shazeer, N.~Parmar, J.~Uszkoreit, L.~Jones, A.~N. Gomez,
  {\L}.~Kaiser, and I.~Polosukhin, ``Attention is all you need,''
  \emph{Advances in Neural Information Processing Systems}, vol.~30, 2017.

\bibitem{gao2023adaptive}
J.~Gao, Y.~Chen, W.~Hu, and D.~Zhang, ``An adaptive deep-learning load
  forecasting framework by integrating transformer and domain knowledge,''
  \emph{Advances in Applied Energy}, vol.~10, p. 100142, 2023.

\bibitem{matar2023transformer}
M.~Matar, P.~G. Estevez, P.~Marchi, F.~Messina, R.~Elmoudi, and S.~Wshah,
  ``Transformer-based deep learning model for forced oscillation
  localization,'' \emph{International Journal of Electrical Power \& Energy
  Systems}, vol. 146, p. 108805, 2023.

\bibitem{zhong2022method}
Z.~Zhong, L.~Guan, Y.~Su, J.~Yu, J.~Huang, and M.~Guo, ``A method of
  multivariate short-term voltage stability assessment based on heterogeneous
  graph attention deep network,'' \emph{International Journal of Electrical
  Power \& Energy Systems}, vol. 136, p. 107648, 2022.

\bibitem{zhang2021deep}
M.~Zhang, J.~Li, Y.~Li, and R.~Xu, ``Deep learning for short-term voltage
  stability assessment of power systems,'' \emph{IEEE Access}, vol.~9, pp.
  29\,711--29\,718, 2021.

\bibitem{zhu2017imbalance}
L.~Zhu, C.~Lu, Z.~Y. Dong, and C.~Hong, ``Imbalance learning machine-based
  power system short-term voltage stability assessment,'' \emph{IEEE
  Transactions on Industrial Informatics}, vol.~13, no.~5, pp. 2533--2543,
  2017.

\bibitem{goodfellow2014generative}
I.~J. Goodfellow, J.~Pouget-Abadie, M.~Mirza, B.~Xu, D.~Warde-Farley, S.~Ozair,
  A.~Courville, and Y.~Bengio, ``Generative adversarial networks,'' \emph{arXiv
  preprint arXiv:1406.2661}, 2014.

\bibitem{arjovsky2017wasserstein}
M.~Arjovsky, S.~Chintala, and L.~Bottou, ``Wasserstein generative adversarial
  networks,'' in \emph{International Conference on Machine Learning}.\hskip 1em
  plus 0.5em minus 0.4em\relax PMLR, 2017, pp. 214--223.

\bibitem{gulrajani2017improved}
I.~Gulrajani, F.~Ahmed, M.~Arjovsky, V.~Dumoulin, and A.~Courville, ``Improved
  training of wasserstein gans,'' \emph{arXiv preprint arXiv:1704.00028}, 2017.

\bibitem{zhu2020time}
L.~Zhu, C.~Lu, and Y.~Luo, ``Time series data-driven batch assessment of power
  system short-term voltage security,'' \emph{IEEE Transactions on Industrial
  Informatics}, vol.~16, no.~12, pp. 7306--7317, 2020.

\bibitem{ren2019fully}
C.~Ren and Y.~Xu, ``A fully data-driven method based on generative adversarial
  networks for power system dynamic security assessment with missing data,''
  \emph{IEEE Transactions on Power Systems}, vol.~34, no.~6, pp. 5044--5052,
  2019.

\bibitem{mai2015semi}
D.~S. Mai and L.~T. Ngo, ``Semi-supervised fuzzy c-means clustering for change
  detection from multispectral satellite image,'' in \emph{2015 IEEE
  International Conference on Fuzzy Systems (FUZZ-IEEE)}.\hskip 1em plus 0.5em
  minus 0.4em\relax IEEE, 2015, pp. 1--8.

\bibitem{li2021privacy}
Y.~Li, J.~Li, and Y.~Wang, ``Privacy-preserving spatiotemporal scenario
  generation of renewable energies: A federated deep generative learning
  approach,'' \emph{IEEE Transactions on Industrial Informatics}, vol.~18,
  no.~4, pp. 2310--2320, 2022.

\bibitem{li2022hierarchical}
Y.~Li, B.~Wang, Z.~Yang, J.~Li, and C.~Chen, ``Hierarchical stochastic
  scheduling of multi-community integrated energy systems in uncertain
  environments via stackelberg game,'' \emph{Applied Energy}, vol. 308, p.
  118392, 2022.

\bibitem{mirza2014conditional}
M.~Mirza and S.~Osindero, ``Conditional generative adversarial nets,''
  \emph{arXiv preprint arXiv:1411.1784}, 2014.

\bibitem{li2023data}
Y.~Li, M.~Han, M.~Shahidehpour, J.~Li, and C.~Long, ``Data-driven
  distributionally robust scheduling of community integrated energy systems
  with uncertain renewable generations considering integrated demand
  response,'' \emph{Applied Energy}, vol. 335, p. 120749, 2023.

\bibitem{chicco2020advantages}
D.~Chicco and G.~Jurman, ``The advantages of the matthews correlation
  coefficient (\uppercase{MCC}) over \uppercase{F}1 score and accuracy in
  binary classification evaluation,'' \emph{BMC Genomics}, vol.~21, no.~1, pp.
  1--13, 2020.

\bibitem{xu2018empirical}
Q.~Xu, G.~Huang, Y.~Yuan, C.~Guo, Y.~Sun, F.~Wu, and K.~Weinberger, ``An
  empirical study on evaluation metrics of generative adversarial networks,''
  \emph{arXiv preprint arXiv:1806.07755}, 2018.

\bibitem{tran2021data}
N.-T. Tran, V.-H. Tran, N.-B. Nguyen, T.-K. Nguyen, and N.-M. Cheung, ``On data
  augmentation for \uppercase{gan} training,'' \emph{IEEE Transactions on Image
  Processing}, vol.~30, pp. 1882--1897, 2021.

\bibitem{van2020test}
T.~Van~Cutsem, M.~Glavic, W.~Rosehart, C.~Canizares, M.~Kanatas, L.~Lima,
  F.~Milano, L.~Papangelis, R.~A. Ramos, J.~A. dos Santos \emph{et~al.}, ``Test
  systems for voltage stability studies,'' \emph{IEEE Transactions on Power
  Systems}, vol.~35, no.~5, pp. 4078--4087, 2020.

\bibitem{bu2015psd}
G.~Bu, Y.~Tang, and J.~Hou, ``\uppercase{PSD-ST} transient stability program
  user manual version 5.0,'' \emph{Beijing: China Electric Power Research
  Institute}, 2015.

\bibitem{kim2020understanding}
J.-K. Kim, B.~Lee, J.~Ma, G.~Verbi{\v{c}}, S.~Nam, and K.~Hur, ``Understanding
  and evaluating systemwide impacts of uncertain parameters in the dynamic load
  model on short-term voltage stability,'' \emph{IEEE Transactions on Power
  Systems}, vol.~36, no.~3, pp. 2093--2102, 2021.

\bibitem{zhang2022load}
X.~Zhang, D.~J. Hill, and Y.~Song, ``A load dynamic stability index for
  short-term voltage stability assessment and control,'' \emph{IEEE
  Transactions on Power Systems}, vol.~38, no.~4, pp. 3304--3316, 2023.

\bibitem{li2023pmu}
Y.~Li, S.~Zhang, Y.~Li, J.~Cao, and S.~Jia, ``\uppercase{PMU}
  measurements-based short-term voltage stability assessment of power systems
  via deep transfer learning,'' \emph{IEEE Transactions on Instrumentation and
  Measurement}, vol.~72, pp. 1--11, 2023.

\bibitem{chawla2002smote}
N.~V. Chawla, K.~W. Bowyer, L.~O. Hall, and W.~P. Kegelmeyer,
  ``\uppercase{SMOTE}: synthetic minority over-sampling technique,''
  \emph{Journal of Artificial Intelligence Research}, vol.~16, pp. 321--357,
  2002.

\bibitem{batista2004study}
G.~E. Batista, R.~C. Prati, and M.~C. Monard, ``A study of the behavior of
  several methods for balancing machine learning training data,'' \emph{ACM
  SIGKDD Explorations Newsletter}, vol.~6, no.~1, pp. 20--29, 2004.

\bibitem{alhudhaif2021novel}
A.~Alhudhaif, ``A novel multi-class imbalanced eeg signals classification based
  on the adaptive synthetic sampling (adasyn) approach,'' \emph{PeerJ Computer
  Science}, vol.~7, p. e523, 2021.

\bibitem{borivcic2021fundamental}
A.~Bori{\v{c}}i{\'c}, J.~L.~R. Torres, and M.~Popov, ``Fundamental study on the
  influence of dynamic load and distributed energy resources on power system
  short-term voltage stability,'' \emph{International Journal of Electrical
  Power \& Energy Systems}, vol. 131, p. 107141, 2021.

\end{thebibliography}
\end{document}